\documentclass[prd,preprint,superscriptaddress]{revtex4}
\usepackage{revsymb}
\usepackage[dvips]{graphicx}

\usepackage{epsfig}
\usepackage{graphicx}
\usepackage{indentfirst}
\usepackage{amsmath}
\usepackage{amsfonts}
\usepackage{amssymb}

\begin{document}

\title{Some remarks on the observational constraints on the\\
self-interacting scalar field model for dark energy}

\author{J\'ulio C. Fabris\footnote{E-mail: fabris@cce.ufes.br} and Deborah F. Jardim\footnote{E-mail: dfjardim@gmail.com}}
\affiliation{Universidade Federal do Esp\'{\i}rito Santo,
Departamento
de F\'{\i}sica\\
Av. Fernando Ferrari, 514, Campus de Goiabeiras, CEP 29075-910,
Vit\'oria, Esp\'{\i}rito Santo, Brazil}

\begin{abstract}
The dark energy component of the cosmic budget is represented by a self-interacting scalar
field. The violation of the null energy condition is allowed.
Hence, such component can also represent a phantom fluid. The model is tested using supernova type
Ia and matter power spectrum data. The supernova test leads to preferred values for
configurations representing
the phantom fluid. The matter power spectrum constraints for the
dark energy equation of state parameter are highly degenerated. 
In both cases, values for the equation of state parameter corresponding to the phantom fluid are highly admitted if no
particular
prior is used.
\newline
\leftline{Pacs: 98.80.-k,95.36.+x}
\end{abstract}
\date{\today}
\maketitle
Observations indicate that the universe today must be in a phase of accelerated expansion
\cite{riess,per,tonry,riessbis,astier}.
In order to explain the evidences for an accelerated universe, keeping the traditional
general relativity theory untouched, it is necessary a fluid with negative pressure, generally called dark energy, since it must not emit any kind of electromagnetic radiation.
At the same time, dark energy must remain a smooth component of the cosmic budget, since
it does not appear in the dynamics of local virialized systems, like galaxy and clusters
of galaxies. This feature
requires also a negative pressure.
\par
Considering a fluid with an equation of state of the type $p = \alpha\rho$, $p$ being the
pressure and $\rho$ the density, the parameter  $\alpha$ must satisfy the condition
$\alpha < - 1/3$ in order to drive the accelerated expansion of the universe remaining
at same time an unclustered component of the cosmic budget: dark energy must
violate the strong energy condition.
\par
There are some claims that the observational data favors negative
values for $\alpha$ such that the null energy condition $p + \rho \geq 0$ ($\alpha < - 1$) is also be violated \cite{caldwell}. If this is the case, the universe may evolve towards a
future singularity, since the violation of the null energy condition may lead
to the divergence of the curvature invariants in a future finite proper time. This is
due to the fact that the violation of the null energy condition implies that
the density of the fluid grows as the universe expands - it is remarkable that
the divergence in the density, and consequently in the curvature, happens in finite proper
time and not asymptoticaly.
\par
It is essential to verify the strength of the evidences for a violation
of the dominant energy condition. A lot of work has been devoted to this
question, see for example \cite{sr} and references there in. The most simple representation of
dark energy is through a fluid with a fixed equation of state of the type
$p = \alpha\rho$, with $\alpha < - 1/3$, the phantom
case corresponding to $\alpha < - 1$. In what concerns the SN Ia test, that is all it is needed as far as
the equation of state of the dark energy component does not evolve. But, for the other tests, like
those requiring perturbative analysis, the real nature of phantom field is crucial, that is, it is essential to know if the results are obtained considering the dark energy component (and in special the phantom fluid) as a self-interacting scalar field or
a scalar field with non-canonical kinetic term, etc. There are many
works exploring different possibilities, see for example references \cite{dutta,ward,cappo}, to quote just
the more recent ones.
In some cases, a phantom phase can be represented in such a way that
it can evolve towards a non-phantom configuration in the future \cite{nasseri,ademir}, leading to the avoidance of the future singularity. This
possibility requires, of course, a non-constant $\alpha$.
\par
In what follows we will consider this simplest case where dark energy, at least at
background level, must satisfy an equation of state of the type
$p = \alpha\rho$, $\alpha$ being a constant. Hence, no evolution of the equation of state parameter $\alpha$
with time will be allowed. Evidently, in such situation, the precise value of $\alpha$ is very
important in order to extrapolate the evolution of the universe for the very distant
future. For example, if $\alpha < - 5/3$, it can appear instabilities in the perturbations
in the very large wavelength limit, inducing an avoidance of big rip due to violation
of the homogeneity and isotropic conditions \cite{fant1,fant2}. Moreover, $\alpha = - 1$ represents the
cosmological constant, and its crucial to understand to which extend it is a
special value in the, otherwise, continous parameter. In considering a constant equation of state for the dark energy fluid, the question
of the future singularity is not addressed. 
\par
In order to do clarify the situation concerning the evidences for a phantom cosmology, at least in the framework considered here,
we will concentrate in two observational tests: the matter power spectrum and
the supernova type Ia data. For the first case, we will consider the data from the
2dFGRS observational program \cite{coles}; for the second, we will use the data of the gold
sample \cite{riessbis}. The model contains two fluids: one pressureless component, which include baryons
and dark matter, and a dark energy component, represented by a fluid which obeys the
equation of state $p = \alpha\rho$, with $\alpha$ in principle negative. The novelty of this analysis consists in the
representation it will be employed for the dark energy fluid, a self-interacting scalar field with a specific potential, and the
absence of any prior in the statistical analysis besides the spatial flatness. Concerning the influence of priors in
the evaluation of cosmological parameters using observational data, in a specific context, see references \cite{liddle,krauss}.
\par
With the two components described above, the Einstein's equations for a flat, isotropic and
homogenous universe described by the
Friedmann-Robertson-Walker metric, reduce to the following expression:
\begin{eqnarray}
 H^2 &=& \Omega_{m0}a^{-3} + \Omega_{x0}a^{-3(1 + \alpha)},
\end{eqnarray}
where $\Omega_{m0} = \Omega_{dm0} + \Omega_{b0}$ is the ratio of matter component to
the critical density today, including
dark matter ($\Omega_{dm0}$) and baryons ($\Omega_{b0}$), while $\Omega_{x0}$ is dark
energy ratio to the critical density.
\par
The SN Ia observational test constrains only the background relation, through the
luminosity distance function. However, the matter power spectrum analysis depends
strongly on the nature of the components. For example, a fluid or a scalar field
representation leads to complete different results for the parameter estimations. The situation is more delicate
when components with negative pressure are considered, as it is the case for dark energy: a fluid description leads
to an imaginary sound velocity, being unstable at small scales, while a field
description through self-interacting scalar field implies positive sound velocity at
sub-horizon scales \cite{nazira}. Since the observational data for the matter power spectrum concerns
sub-horizon modes, the specific descriptions for dark energy and dark matter are fundamental
to interpret the observational constraints.
\par
To cope with the instability problem described above (which to some extend excludes the possibility of an ordinary fluid description
for the dark energy component), the dark energy field will be
described through a self-interacting scalar field. This is the
simplest field description of a given component
in cosmology. If it is a realistic description or a unique one (certainly, to some extent, it is not) is outside
the aims of the present work: it would require, to answer this question, to know
the origin of the dark energy field, what is object of speculation, with no clear
candidate. A comparison of specific model with the observational data aids, of course, to shed some light on this
question.
\par
In the absence of matter field, it is quite easy to reproduce the behaviour of a dark energy field
through a self-interacting
scalar field. The Friedmann equation coupled to
a self-interacting scalar field and the Klein-Gordon equation for the scalar field reads,
\begin{eqnarray}
 3\biggr(\frac{a'}{a}\biggl)^2 = \epsilon\frac{\phi'^2}{2} + V(\phi)a^2, \\
\phi'' + 3\frac{a'}{a}\phi' = - \epsilon\frac{dV(\phi)}{d\phi}a^2,
\end{eqnarray}
where the primes mean derivative with respect to the conformal time $\eta$ defined
by the expression $dt = a(\eta)d\eta$, and $\epsilon = + 1$ is required to describe a "normal" dark energy fluid,
while for a phantom fluid $\epsilon = - 1$: for a phantom field, the kinetic term must have the ``wrong'' sign.
For a general equation of state $p = \alpha\rho$, the scale factor
behaves as $a = a_0\eta^{2/(1 + 3\alpha)}$. This behaviour can
be reproduced by a self-interacting scalar field with the form \cite{fant1,fant2},
\begin{equation}
\label{pot1}
 V(\phi) = V_0\,e^{\pm\sqrt{3\epsilon(1 + \alpha)}\phi}\quad , \quad \phi(\eta) = \pm2\frac{\sqrt{3\epsilon(1 + \alpha)}}{1 + 3\alpha}\ln\eta \quad ,
\end{equation}
$V_0$ being a constant. It is not a surprise the appearence of an exponential potential, see for example reference \cite{kujat}.
An interesting analysis for a power law potential, implying a non-constant equation of state, has been performed in reference \cite{siri}.
\par
When matter is present, the potential (\ref{pot1}) does not represent anymore exactly
the dynamics of the dark energy fluid. In fact, this representation is exact only in the 
asymptotic limit. When pressureless matter is present the potential that reproduces
the coupled system dark energy/matter is more complicated, and it can not be represented, apparently, in
a closed form using elementary functions. However, the scalar field and the potential
for this case can be implicitly expressed in terms of the scale factor.
The overall dynamics is accounted by the following expressions:
\begin{eqnarray}
 \biggr(\frac{a'}{a}\biggl)^2 &=& \Omega_{m0}a^{-3} + \Omega_{x0}a^{-3(1 + \alpha)},\\
\Omega_{x} &=& \Omega_{x0}a^{-3(1 + \alpha)} = \epsilon\frac{{\phi'}^2}{2} + Va^2, \\
\phi' &=& \sqrt{3\Omega_{x0}}\sqrt{\epsilon(1 + \alpha)}a^{-3(1 + \alpha)/2} \quad , \quad 
V = \frac{3}{2}(1 - \alpha)a^{- 3(1 + \alpha)}.
\end{eqnarray}
The relation between the scalar field and its potential to the scale factor is
obtained by imposing
that
\begin{eqnarray}
 \epsilon\frac{\phi'^2}{2} + V\,a^2 &=& 8\pi G\rho_x = 8\pi G\rho_{x0}a^{- (1 + 3\alpha)},\\
 \epsilon\frac{\phi'^2}{2} - V\,a^2 &=& 8\pi G p_x = 8\pi G\alpha\rho_{x0}a^{- (1 + 3\alpha)}.
\end{eqnarray}
When $\Omega_{m0} = 0$ these expressions can be re-inserted in the 
Einstein's equation in order to have the explicit dependence of the potential in
terms of the scalar field $\phi$.
The cosmological constant case is reproduced in the sense that the kinetic term becomes
zero for $\alpha = - 1$, and we remain only with a constant potential term.
\par
The Supernova type Ia analysis can be performed by using the moduli distance quantity
defined by
\begin{equation}
 \mu = 5\log_{10}(D_l/Mpc) + 25 ,
\end{equation}
where the luminosity distance $D_L$ is given by
\begin{equation}
 D_L = (1 + z)\frac{c}{H_0}\int_0^z\frac{dz'}{\sqrt{\Omega_{m0}(1 + z')^3 + (1-\Omega_{m0})(1 + z')^{3(1 + \alpha)}}} \quad,
\end{equation}
where $z$ is the redshift.
This expression is valid for a flat universe for which $\Omega_{m0} + \Omega_{x0} = 1$.
We parametrize the Hubble parameter today writing $H_0 = 100\,h\,km/(Mpc.s)$. The model contains three free parameters, $\alpha$, $\Omega_{dm0}$ and $h$, while the baryonic component is fixed such that $\Omega_{b0} = 0.04$. We use the SN Ia gold sample. The $\chi^2$ statistics is defined by
\begin{equation}
 \chi^2_{SN} = \sum_{i=1}^N\frac{(\mu_i^t - \mu_i^o)^2}{\sigma_i^2},
\end{equation}
where $\mu_i^o$ is the observational data for the moduli distance for
the $i^{th}$ supernova, $\mu_i^t$ the corresponding theoretical prediction
and $\sigma_i^2$ is the observational bar error including the dispersion velocity.
The probability distribution function (PDF) is obtained through the expression
\begin{equation}
 P(h,\Omega_{m0},\alpha) = Ae^{-\chi^2_{SN}/2},
\end{equation}
where $A$ is a normalization constant.
The PDF is three dimensional. Two dimensional and one dimensional PDF can be obtained integrating in one or two variables.
\par
Let us now turn to the perturbative analysis. We will use the Bardeen's gauge-invariant formalism. For the case including pressureless matter and a self-interacting scalar field,
the perturbed equations read (see \cite{mukhanov1}):
\begin{eqnarray}
 \nabla^2\Phi - 3{\cal H}\Phi' - \biggr[3{\cal H}^2 - \epsilon\frac{\phi'^2}{2}\biggl]\Phi &=& 4\pi Ga^2\delta\rho + \epsilon\frac{\phi'}{2}\delta\phi' + \frac{V_\phi}{2}a^2\delta\phi, \\
\Phi'' + 3{\cal H}\Phi' + \biggr[2{\cal H}' + {\cal H}^2 + \epsilon\frac{\phi'^2}{2}\biggl]\Phi &=& 4\pi Ga^2\delta p + \epsilon\frac{\phi'}{2}\delta\phi'
- \frac{V_\phi}{2}a^2\delta\phi, \\
\delta\phi'' + 2{\cal H}\delta\phi' - \nabla^2\delta\phi + \epsilon V_{\phi\phi}a^2\delta\phi &=& 4\phi'\Phi'
- 2\epsilon V_{\phi}a^2\Phi.
\end{eqnarray}
In these expressions, ${\cal H} = a'/a$, and the subscript $\phi$ indicate derivative with respect to $\phi$. Since the fluid represents matter,
$\delta p = 0$. The anisotropic pressure is made equal to zero in these equations.
\par
It is convenient to use the scale factor $a$ as the new variable. The perturbed equations
and the background relations can be re-expressed in terms of this new variable.
For the perturbed equations we find:
\begin{eqnarray}
\label{pe1}
 \ddot\Phi + \biggr[\frac{3}{a} + \frac{a''}{a'^2}\biggl]\dot\Phi 
+ \biggr[2\frac{a''}{aa'^2} - \frac{1}{a^2} + \epsilon\frac{\phi'^2}{2a'^2}\biggl]\Phi &=&
\epsilon\frac{1}{2}\frac{\phi'}{a'}\dot\lambda - \frac{V_\phi}{2}\frac{a^2}{a'^2}\lambda, \\
\label{pe2}
\ddot\lambda + \biggr[\frac{2}{a} + \frac{a''}{a'^2}\biggl]\dot\lambda +
\biggr\{\biggr(\frac{k\,l_0}{a'}\biggl)^2 + \epsilon V_{\phi\phi}\frac{a^2}{a'^2}\biggl\}\lambda &=& 4\frac{\phi'}{a'}\dot\Phi - 2\epsilon V_\phi\frac{a^2}{a'^2}\Phi,
\end{eqnarray}
where $\lambda = \delta\phi$ and the dots mean now derivative with respect to
$a$.
We have the following definitions:
\begin{eqnarray}
 a' &=& \sqrt{\Omega_{m0}a + \Omega_{x0}a^{(1 - 3\alpha)}},\\
a'' &=& \frac{1}{2}[\Omega_{m0} + (1 - 3\alpha)\Omega_{x0}a^{- 3\alpha}],\\
\phi' &=& \sqrt{3|1 +\alpha|\Omega_{x0}}a^{-3(1 + \alpha)/2},\\
V(a) &=& \frac{3}{2}\Omega_{x0}(1 - \alpha)a^{- 3(1 + \alpha)},\\
V_\phi(a) &=& - \frac{3}{2}(1 - \alpha)\sqrt{3\Omega_{x0}|1 + \alpha|}
a^{-(7 + 3\alpha)/2}a',\\
V_{\phi\phi}(a) &=& \frac{a'}{\phi'}\frac{d}{da}V_\phi(a),
\end{eqnarray}
where the subscript $\phi$ means derivative with respect to the scalar field.
Moreover, $k$ is the wavenumber of the perturbation coming from
the Fourier decomposition and $l_0 = 3.000\cdot h\, Mpc$  is the Hubble radius today.
\par
We will perform a numerical integration of equations (\ref{pe1},\ref{pe2}). The initial
conditions are used employing the BBKS transfer function, and supposing a Harrison-Zeldovich
primordial spectrum \cite{martin,sugiyama,bardeen}. The implementation of the initial conditions is described in reference
\cite{sola}.
We will compute the matter power spectrum, defined as,
\begin{equation}
 P_k = |\delta_k|^2,
\end{equation}
$\delta_k$ being the Fourier component of the density contrast.
As in the preceding SN case, we can evaluate the $\chi^2$ parameter that gives the quality of the fitting of
the observational data by the theoretical model:
\begin{equation}
 \chi^2_{PS} = \sum_{i=1}^N\frac{(P_{k_i}^{th} - P_{k_i}^{ob})^2}{\sigma_i^2},
\end{equation}
where $k_i$ corresponds to the $i^{th}$ Fourier mode, $P_{k_i}^{th}$ is the theoretical
prediction for this mode, $P_{k_i}^{ob}$ is the corresponding observational data, and
$\sigma_i$ its observational uncertainty.
Since we use modes corresponding to the linear regime (scales larger than $10\,Mpc$),
it is not necessary to use the full correlation matrix.
\par
From the $\chi^2_{PS}$, we can define the probability density function (PDF) as
\begin{equation}
 P(\Omega_{dm0},\alpha) = Ae^{-\chi^2_{PS}/2},
\end{equation}
where $A$ is a normalization factor. It depends, as indicated, on two free parameters, the
dark matter fractional density $\Omega_{dm0}$ and on the equation of state parameter,
$\alpha$. The baryonic density is fixed as before.
\par
\begin{center}
\begin{figure}[!t]
\begin{minipage}[t]{0.3\linewidth}
\includegraphics[width=\linewidth]{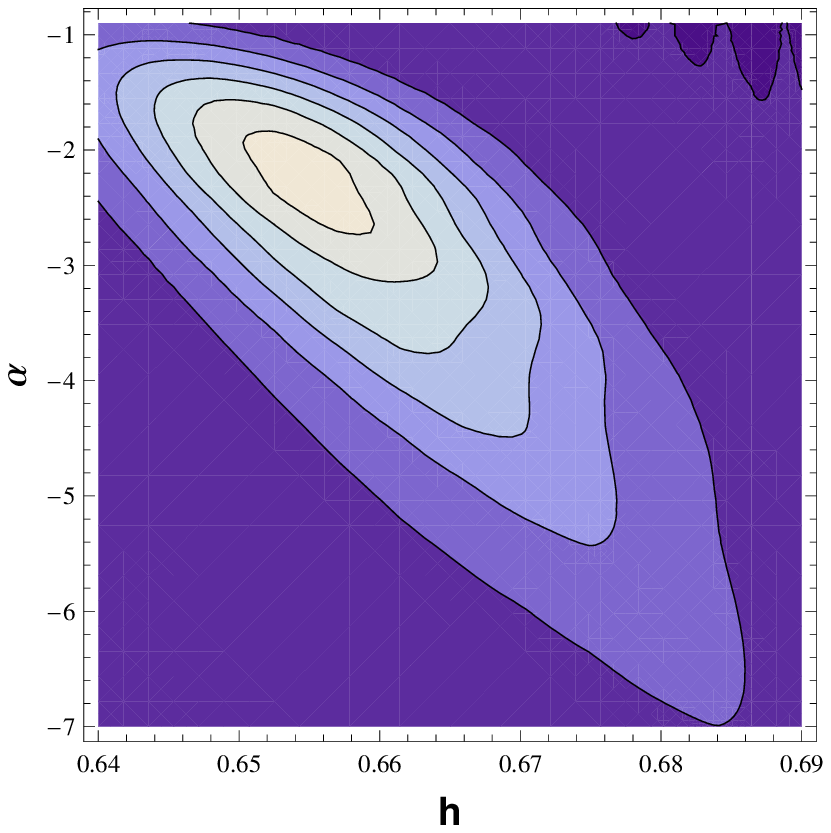}
\end{minipage} \hfill
\begin{minipage}[t]{0.3\linewidth}
\includegraphics[width=\linewidth]{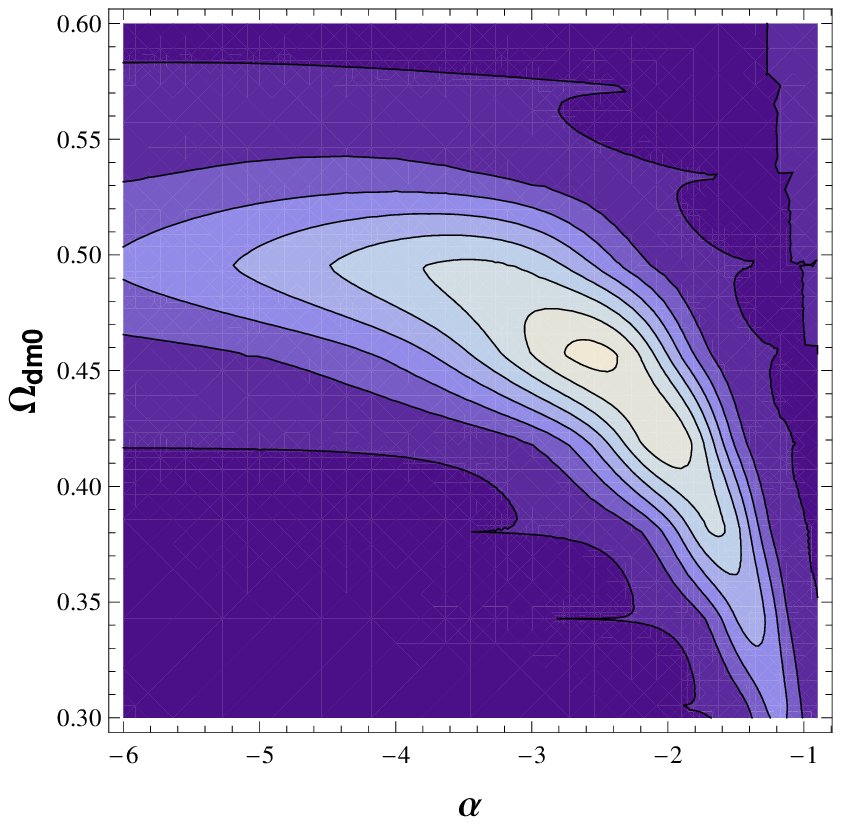}
\end{minipage} \hfill
\begin{minipage}[t]{0.3\linewidth}
\includegraphics[width=\linewidth]{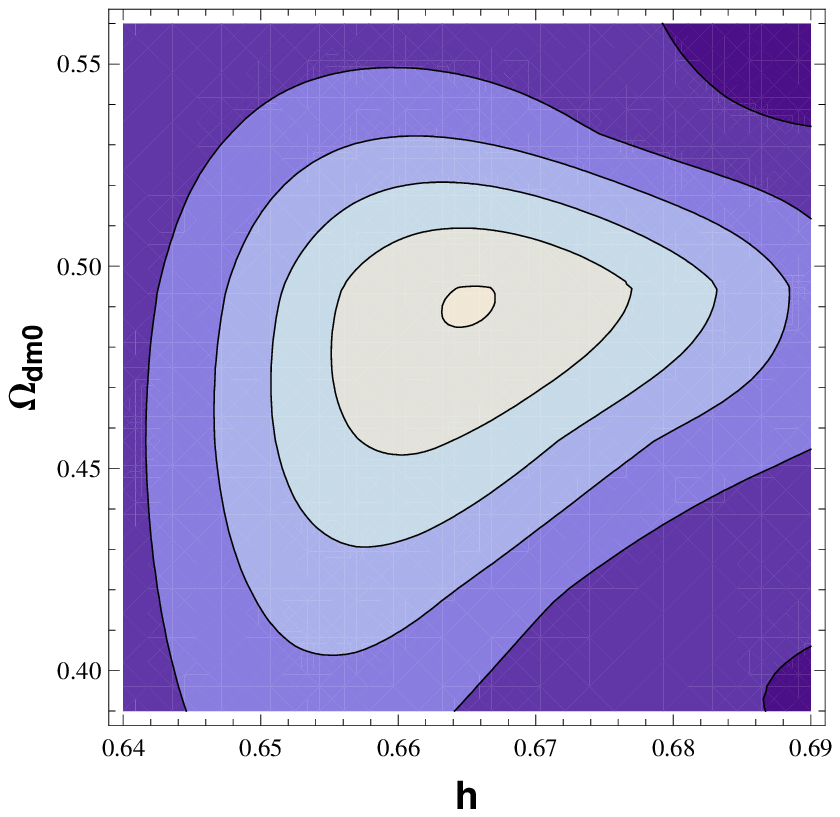}
\end{minipage} \hfill
\begin{minipage}[t]{0.3\linewidth}
\includegraphics[width=\linewidth]{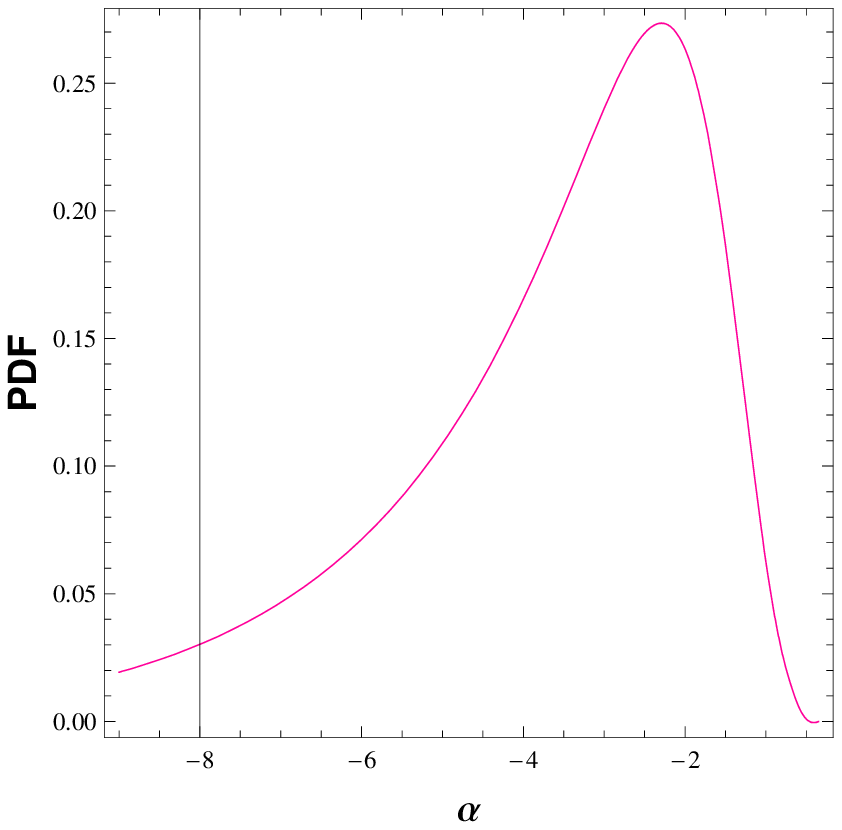}
\end{minipage} \hfill
\begin{minipage}[t]{0.3\linewidth}
\includegraphics[width=\linewidth]{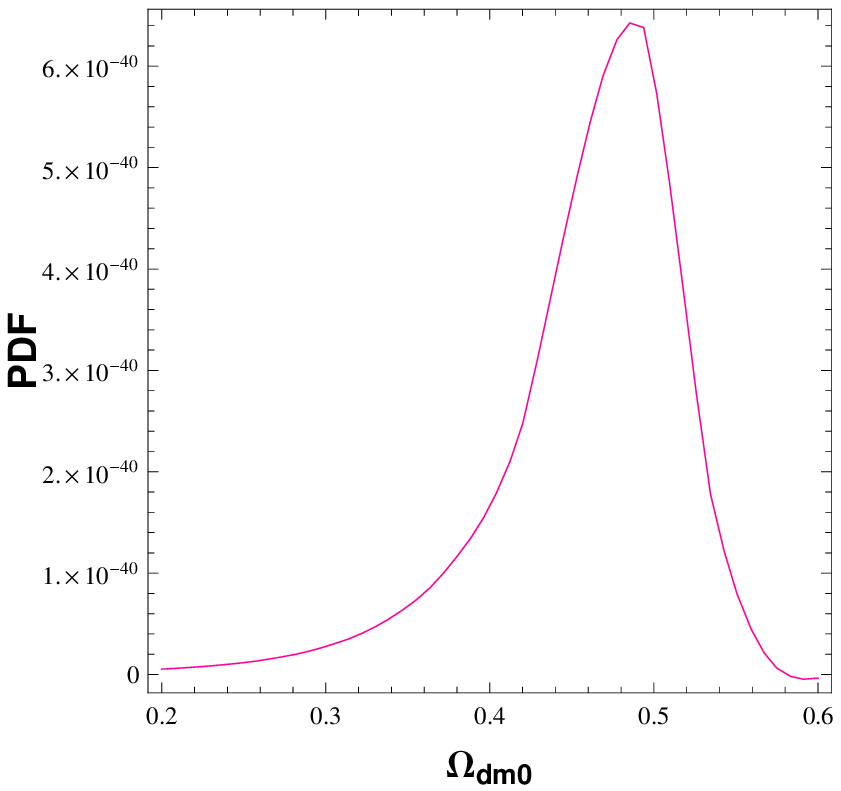}
\end{minipage} \hfill
\begin{minipage}[t]{0.3\linewidth}
\includegraphics[width=\linewidth]{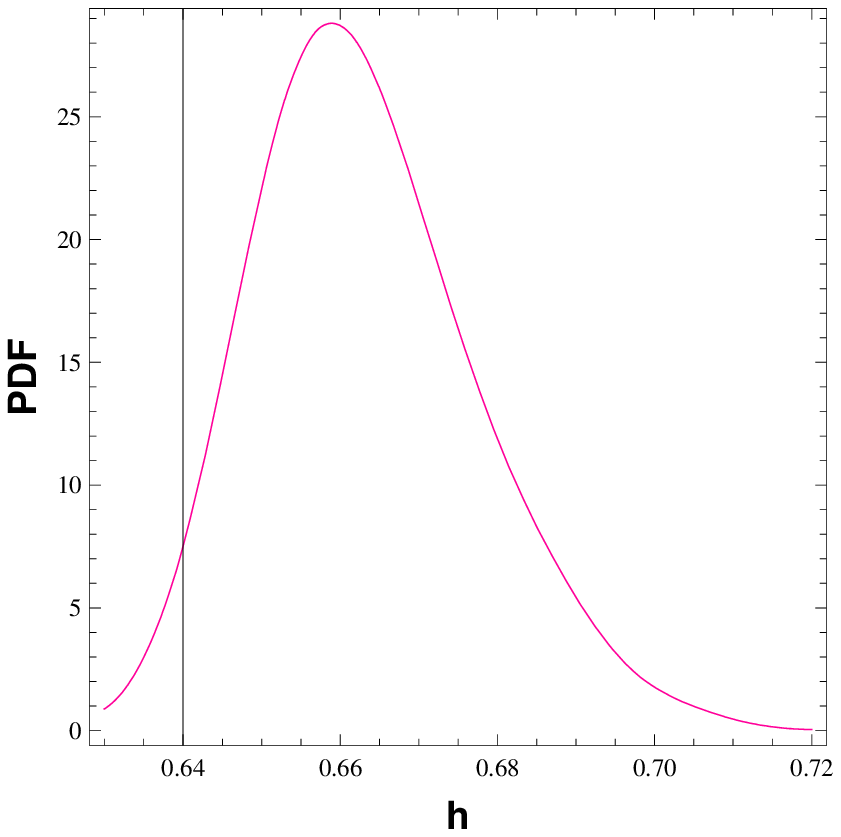}
\end{minipage} \hfill
\caption{{\protect\footnotesize The two-dimensional PDF using the SN Ia data for different combinations
of $h$, $\Omega_{dm0}$ and $\alpha$ are shown in the top pannels. In the bottom panels, it is displayed
the corresponding one-dimensional PDF.}}
\label{fig1}
\end{figure}
\end{center}

In figure \ref{fig1} we display the two and one dimensional PDF when only the SN Ia data (gold sample)
are used. As explained before, since the spatial section is supposed flat, there are
three independent parameters: $h$, $\Omega_{dm0}$ and $\alpha$.
Minimizing $\chi^2_{SN}$, we find $\Omega_{dm0} = 0.47$, $\alpha = - 2.40$, $h = 0.66$, with 
$\chi^2_{SN} = 1.11$. The two (one) dimensional
PDF is obtained integrating on the remaining one (two) parameters.
The two-dimensional probability distributions show that values around
$\alpha = - 2$ and $h = 0.65$ are favored. This is confirmed after marginalization: the
peaks of probability occur at $\alpha = - 2.29$ and $h = 0.66$. Remark that the probability
for $\alpha$ decreases after the maximum, but slowly. The extension of the range of $\alpha$ to very large negative values has
increased a little the estimated value of $h$, in the direction of the value predicted by the CMB test which is around $h = 0.72$.
These results agrees in their general lines with those of reference \cite{sr}, where the origin of the dark energy
component is traced back to quantum effects, and the constraints are obtained by imposing that
the resulting scalar field does not have a mass larger than the Planck mass: in this work, for the configuration
corresponding to our model, there is a maximum of probability around $\alpha = - 2$.
For $\Omega_{dm0}$ the analysis is a little more delicate. The one-dimensional PDF predicts
a peak at $\Omega_{m0} = 0.49$, a large value compared
to the $\Lambda$CDM model, for which $\Omega_{dm0} \sim 0.25$ \cite{tonry,riessbis}. But, remark that, from the
two-dimensional PDF for $\alpha$ and $\Omega_{dm0}$, it is clear that large negative values
for $\alpha$ demands larger values for $\Omega_{dm0}$. Hence, after normalization, extending
the integration to deep negative values of $\alpha$ increases the predicted
value for $\Omega_{dm0}$.

\begin{center}
\begin{figure}[!t]
\begin{minipage}[t]{0.3\linewidth}
\includegraphics[width=\linewidth]{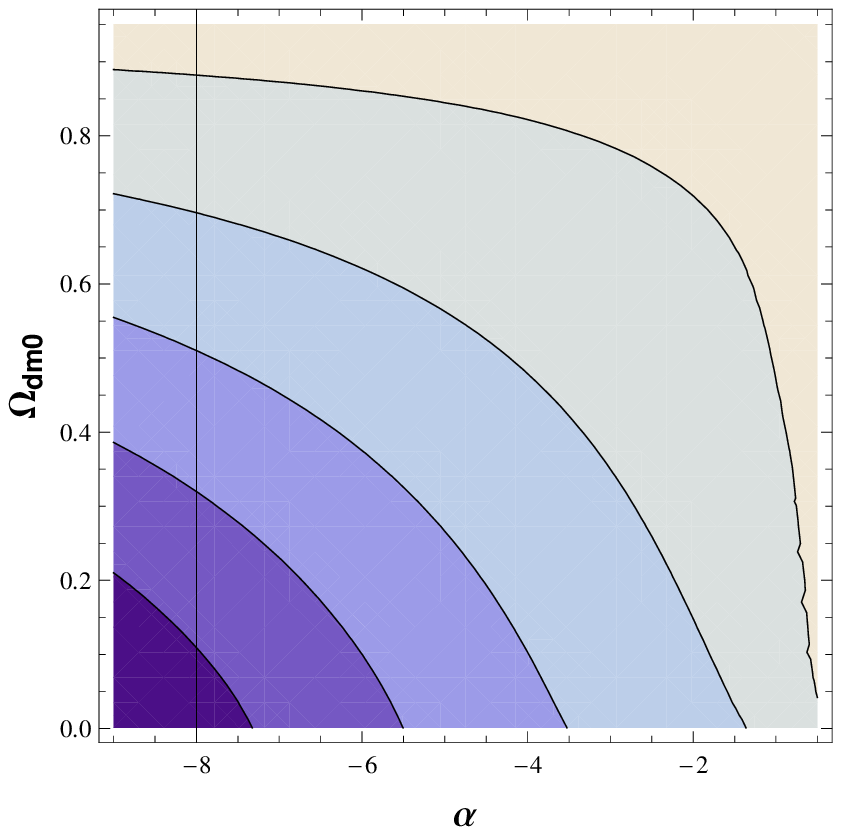}
\end{minipage} \hfill
\begin{minipage}[t]{0.3\linewidth}
\includegraphics[width=\linewidth]{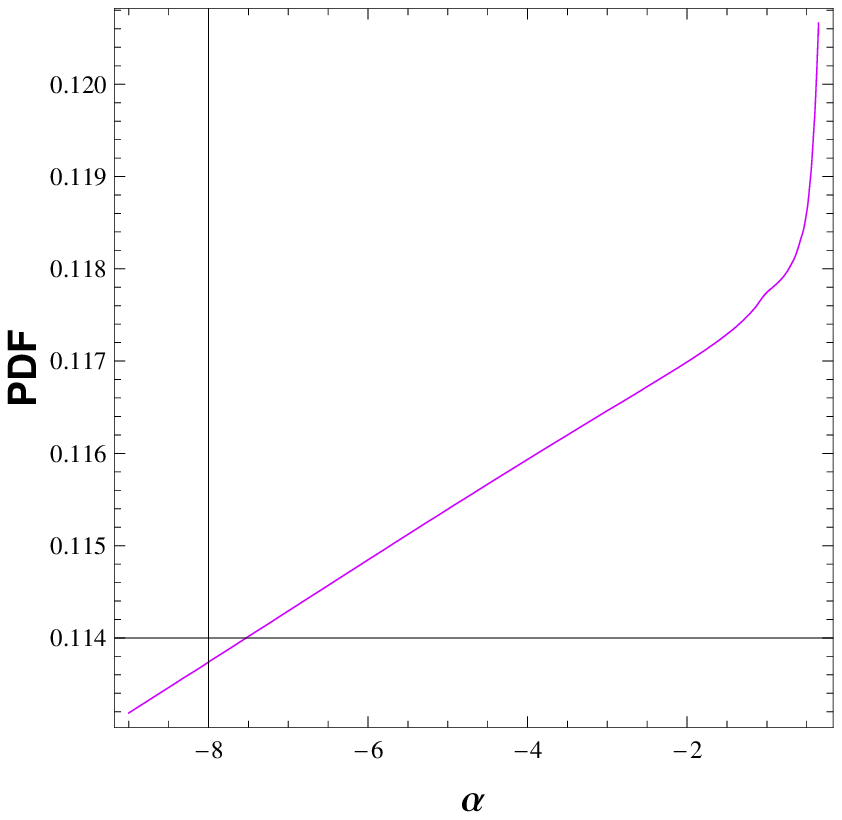}
\end{minipage} \hfill
\begin{minipage}[t]{0.3\linewidth}
\includegraphics[width=\linewidth]{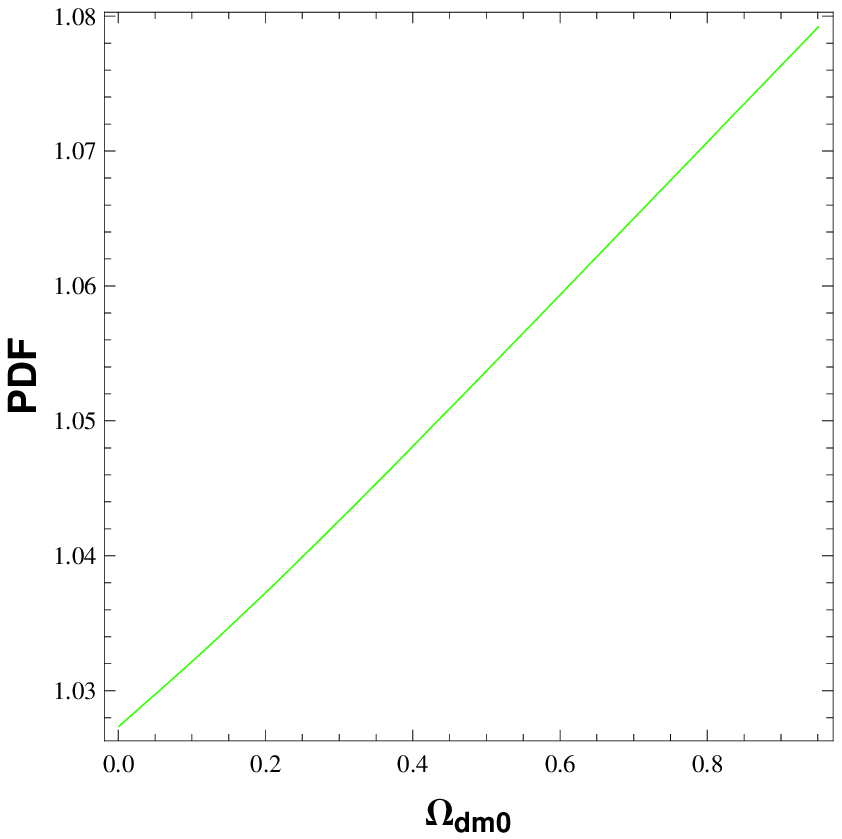}
\end{minipage} \hfill
\caption{{\protect\footnotesize The two-dimensional PDF using the matter power
spectrum data for $\Omega_{dm0}$ and $\alpha$ is shown in the left pannels. In the 
center and right panels, it is displayed
the corresponding one-dimensional PDF. Remark that the probability is almost constant.}}
\label{fig2}
\end{figure}
\end{center}

\begin{center}
\begin{figure}[!t]
\begin{minipage}[t]{0.3\linewidth}
\includegraphics[width=\linewidth]{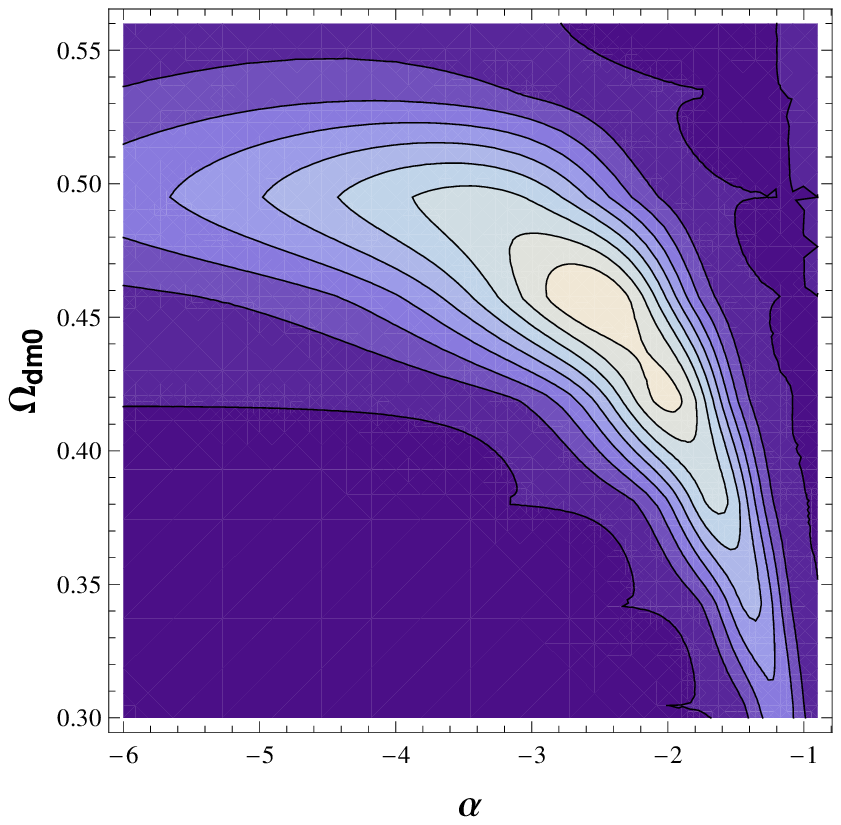}
\end{minipage} \hfill
\begin{minipage}[t]{0.3\linewidth}
\includegraphics[width=\linewidth]{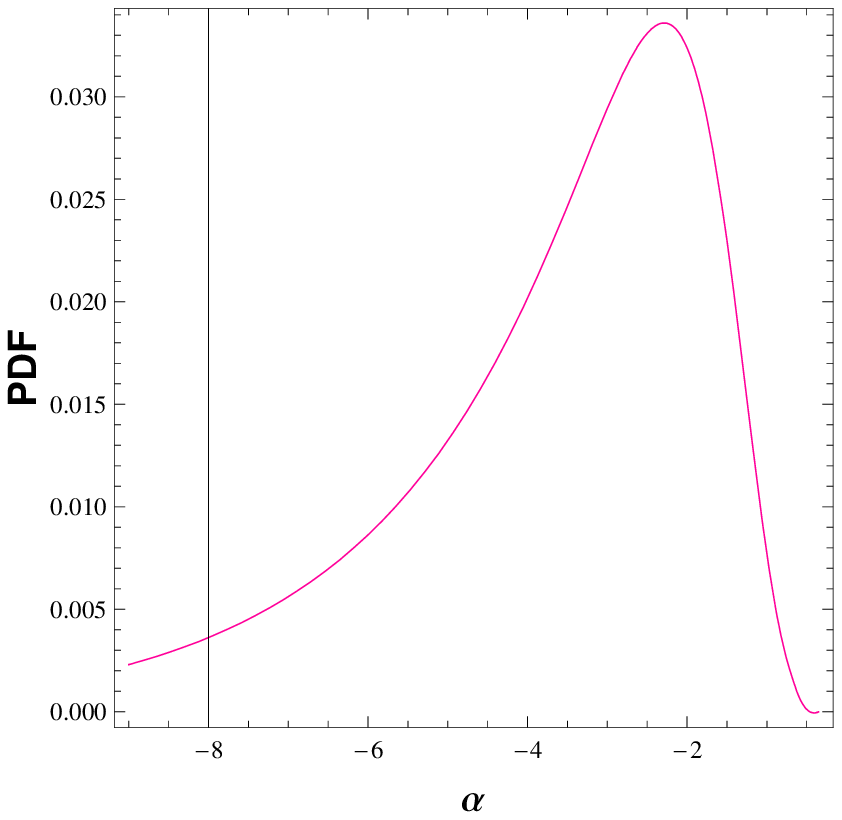}
\end{minipage} \hfill
\begin{minipage}[t]{0.3\linewidth}
\includegraphics[width=\linewidth]{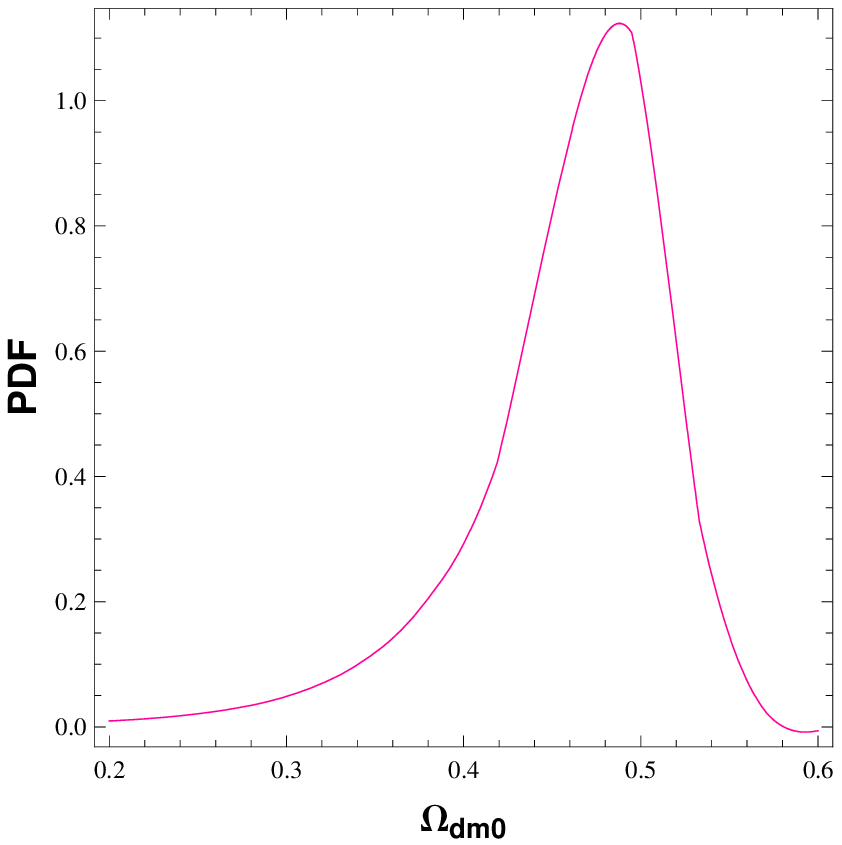}
\end{minipage} \hfill
\caption{{\protect\footnotesize The two-dimensional PDF for $\Omega_{dm0}$ and $\alpha$ using both the SN Ia and the matter power
spectrum data is shown in the left pannels. In the 
center and right panels, it is displayed
the corresponding one-dimensional PDF.}}
\label{fig3}
\end{figure}
\end{center}

\par
The power spectrum analysis put new constraints. The power spectrum analysis constrains
$\Omega_{dm0}$ and $\alpha$, the results being function of $h$. The best fit scenario implies
$\alpha = - 0.90$, $\Omega_{dm0} = 0.95$ with $\chi^2_{PS} = 0.38$. The two and one dimensional PDF are shown in figure \ref{fig2}.
It is clear that positive values for $\alpha$ are excluded, and after
$\alpha \sim - 1/3$ a plateau is reached. Nothing special seems to happen for $\alpha = - 5/3$, the new critical point
identified in references \cite{fant1,fant2}, but we remember that
we have evaluated the power spectrum for the matter component. Hence, there is no contradiction with the results of 
reference \cite{fant1,fant2}.
After marginalization, the probability is essentially constant from $\alpha \sim - 0.3$ on; for $\Omega_{dm0}$ it occurs near
$1$. The variation in the PDF is very small, for both parameter, as far as $\alpha < - 0.3$. 
\par
Composing the joint PDF for both set of data we obtain the two and one-dimensional
PDF displayed in figure \ref{fig3}. The maximum PDF for $\Omega_{dm0}$ is  again at $0.49$, the same position as in the SN Ia case. The maximum
PDF for the equation of state parameter is at $\alpha = - 2.29$ as in the pure SN Ia case.
\par
The analysis here is restricted to the case where the dark energy component is described
by a self-interacting scalar field leading to a constant equation of state. In this model, the evolution with
a constant equation of state corresponds to a critical point in the phase space,
but it is not the only possibility. Introducing perturbations, the effective equation
of state changes, and that is why instabilities do not appear even when $\alpha$ is
negative. This is convenient in order to perform the power spectrum analysis. 
The main message encoded in the results obtained here, concerning the self-interacting scalar field model for dark energy analysed in this
work, seems to be the following: there are strong evidences for a phantom fluid with a very negative value for the equation of state
parameter $\alpha$, mainly due to the SN Ia constraint; otherwise, using
only matter power spectrum, the only clear restriction is that $\alpha$ must be smaller than $\alpha \sim - 1/3$.
It is important to stress that no special prior has been used, in opposition with the analysis made, for example, in references
\cite{tonry,riessbis}. If we particularize the value of the dark energy component for that used in the prior of \cite{tonry,riessbis}
we find essentially their results, with a peak in the probability distribution for $\alpha$ around $ - 1$.
\par
\noindent
{\bf Acknowledgments:} We thank CNPq (Brazil) and FAPES (Brazil) for partial financial support. We thank the kind
hospitality of the {\it Institut d'Astrophysique de Paris - IAP} during part of the elaboration of this work. We thank
also Winfried Zimadahl and Hermano Velten for their suggestions and comments.

\end{document}